\documentclass[%
 reprint,
nofootinbib,
longbibliography,
 amsmath,amssymb,
 aps,
]{revtex4-1}

\usepackage{graphicx}
\usepackage{dcolumn}
\usepackage{bm}
\usepackage{xspace}
\usepackage{multirow}
\usepackage{xcolor}

\usepackage{units}
\usepackage{comment}

\def\sib{{\textsc{Sibyll}\,2.3c}\xspace}
\def\sibColor{yellow}

\def\qgs{\textsc{QGSJet}-II.04\xspace}
\def\qgsColor{red}
\def\epos{\textsc{EPOS-LHC}\,\xspace}
\def\eposColor{green}

\def\Nmu{$N_\mu$\xspace}%
\def\LogNmu{$\ln  N_\mu$\xspace}%
\def\Lmu{$\Lambda_\mu$\xspace}%
\def\Lalpha{$\Lambda_\alpha$\xspace}%
\def\Lpi{$\Lambda_\pi$\xspace}%

\newcommand{\figsize}{0.90}

\begin{document}

\title{Constraining the energy spectrum of neutral pions in ultra--high--energy proton--air interactions}

\author{Lorenzo Cazon}
\address{Laborat\'{o}rio de Instrumenta\c{c}\~{a}o e F\'{i}sica Experimental de Part\'{i}culas (LIP) - Lisbon, Av.\ Prof.\ Gama Pinto 2, 1649-003 Lisbon, Portugal}

\author{Ruben Concei\c{c}\~{a}o}
\address{Laborat\'{o}rio de Instrumenta\c{c}\~{a}o e F\'{i}sica Experimental de Part\'{i}culas (LIP) - Lisbon, Av.\ Prof.\ Gama Pinto 2, 1649-003 Lisbon, Portugal}
\address{Instituto Superior T\'ecnico (IST), Universidade de Lisboa, Av.\ Rovisco Pais 1, 1049-001 Lisbon, Portugal}

\author{Miguel Alexandre Martins}
\address{Laborat\'{o}rio de Instrumenta\c{c}\~{a}o e F\'{i}sica Experimental de Part\'{i}culas (LIP) - Lisbon, Av.\ Prof.\ Gama Pinto 2, 1649-003 Lisbon, Portugal}
\address{Instituto Superior T\'ecnico (IST), Universidade de Lisboa, Av.\ Rovisco Pais 1, 1049-001 Lisbon, Portugal}

\author{Felix Riehn}
\email{friehn@lip.pt}
\address{Instituto Galego de F\'isica de Altas Enerx\'ias (IGFAE),
Universidade de Santiago de Compostela, 15782 Santiago de Compostela, Spain}
\address{Laborat\'{o}rio de Instrumenta\c{c}\~{a}o e F\'{i}sica Experimental de Part\'{i}culas (LIP) - Lisbon, Av.\ Prof.\ Gama Pinto 2, 1649-003 Lisbon, Portugal}

\date{\today}

\begin{abstract}
  
Fluctuations in the muon content of extensive air showers are
anticorrelated to the fluctuations of the energy taken by the neutral
pions which emerge from the first interaction of the cosmic ray in
the atmosphere. We show that the high-energy tail of the neutral
pion spectrum produced in the first proton-air interaction can be
constrained, within the uncertainties of present cosmic ray experiments,
through the analysis of the shower-to-shower distribution of the muon content of the air showers,~$P(N_{\mu})$.

\end{abstract}

\pacs{Valid PACS appear here}
\maketitle


\section{\label{sec:intro}Introduction}
Ultra-high-energy cosmic rays (UHECRs) have long been seen as a unique opportunity to probe hadronic interaction physics at center-of-mass energies that surpass the $100\,$TeV scale, which is well beyond the reach of any human-made collider. Indispensable to this quest is the knowledge of the UHECRs composition, whose average mass number has been shown to be heavier than proton at the highest energies~\cite{AugerXmaxMass}. This result could be explained by phenomena in the physics of UHECR propagation and astrophysical sources that have not been accounted for~\cite{UHECR_review}. Alternatively, it could be due to an incomplete description of hadronic interactions, which might hamper the interpretation of air shower data in terms of the nature of the primaries.

The description of hadronic interactions in air showers is mostly based on phenomenological models. These models are tuned to accelerator data and then extrapolated to energies and kinematic regions essential to the description of the development of the extensive air shower~(EAS). In fact, there are several pieces of evidence showing that the hadronic component of the shower is poorly described. One example is the measurements of the average muon number in showers.

 The combination of several experiments has shown that the muon deficit in simulations starts around $\sim 10^{16}$ eV and steadily increases up to the highest energies available  $\sim 10^{20}\,$eV~\cite{WHISP}.
The origin and nature of this discrepancy are still unknown. In particular, it is unclear if it is related to a poor description of low energy interactions or due to unexpected new phenomena at the highest energies.

Recently, the relative fluctuations and the event-by-event distribution of the number of muons in showers at the highest energies were measured for the first time~\cite{RMS:ICRC2019}.

It was also shown that the fluctuations can be traced back to fluctuations of the quantity $\alpha_1$, which is related to the first interaction of the UHECR~\cite{PLBalpha}.
The quantity $\alpha_1$ is defined as 
\begin{equation}
  \alpha_1 \equiv \sum_{i}^m \left( \frac{E_i^{\rm had}}{E_0}\right)^\beta , \
  \label{eq:alpha1}
\end{equation}
where we sum over $m$ \emph{hadronically interacting} particles (basically all baryons, kaons and pions, excluding neutral pions) and  $E_i^{\rm had}/E_0$ denotes the fraction of the energy of the primary carried by hadron $i$. The parameter $\beta$ is set to $0.93$, motivated by the dependence of the average number of muons with the primary energy in the models. 
In many practical cases, $\alpha_1$ can be approximated by the energy fraction carried by hadronically-interacting particles, and its complement $f_{\textrm{e.m.}}\simeq 1-\alpha_1$ can be taken to be the fraction of energy taken by the $\pi^0$.
Therefore, the total number of muons in the shower is
\begin{equation}
    N_\mu \propto \alpha_1 \alpha_2 \dots\alpha_g\dots\alpha_n
    \label{eq:nmu}
\end{equation}
where the index $g$ runs over generations\footnote{Particle reactions within a cascade can be grouped in generations: the first generation, $g=1$, contains solely the first interaction between the cosmic ray and the air nucleus. The second generation, $g=2$, contains all the first interactions of the secondaries after $g=1$, and so on so forth.} until the cascade stops at generation $g=n$. For the calculation of the second generation, there are $m$ particle interactions contributing to $\alpha_2$, therefore the fluctuations of each single reaction are statistically likely to be compensated by another. The distribution of $\alpha_2$ is narrower by a factor $1/\sqrt{m}$. The same applies for deeper generations $g$: the corresponding distribution of $\alpha_g$ get even narrower as the number of particles in the cascade exponentially increases.
As a consequence, a high degree of correlation between the number of muons and the variable $\alpha_1$ is reached.

On the other hand, for the average number of muons, all generations have the same importance and can equally contribute to explain the muon deficit in simulations. For instance, a small deviation of around 5\% in $\alpha_g$ from $g=1$ up to $g=6$ converts into 30\% when inserted in equation~\eqref{eq:nmu} as $(1+0.05)^6\simeq1.3$.

The detailed analysis of the shower-to-shower distribution of the muon content opens a new window of observation, which brings information on the hadronic interactions at the start of the air shower.

\section{\label{sec:alpha}Exploring showers with low muon content}

This work is based on air shower simulations done with CONEX~\cite{CONEX}, which combine Monte Carlo simulations with cascade equations for low energy particles. This hybrid technique allows a significant increase in the size of the shower samples while preserving the features of the muon distributions.

The $\alpha$-distribution of a proton-Air interaction displays a prominent exponential long tail toward low $N_\mu$ values. The second and third generations suppress this tail by the self-convolution of the individual $\alpha$ distributions of each particle reaction. As shown in~\cite{PLBalpha}, the complementary distribution of the shower, $\omega=\alpha_2 \alpha_3... \alpha_n$, which contains all interactions but the first, shows no long or exponential tail.

In CONEX simulations, the interactions at the highest energies are individually simulated and in full, and the effect of low energy interactions is taken into account by solving cascade equations. The resulting simulations contain all the physics necessary to accurately describe the distribution of $N_\mu$, especially at low values. The most extreme low $N_\mu$ fluctuations can only be explained by a large fraction of electromagnetic (EM) energy in the first interaction and never by the coincidence of a very large number of low energy interactions with large EM fractions.

\begin{figure}[!t]
  \centering
  \includegraphics[width=\figsize\columnwidth]{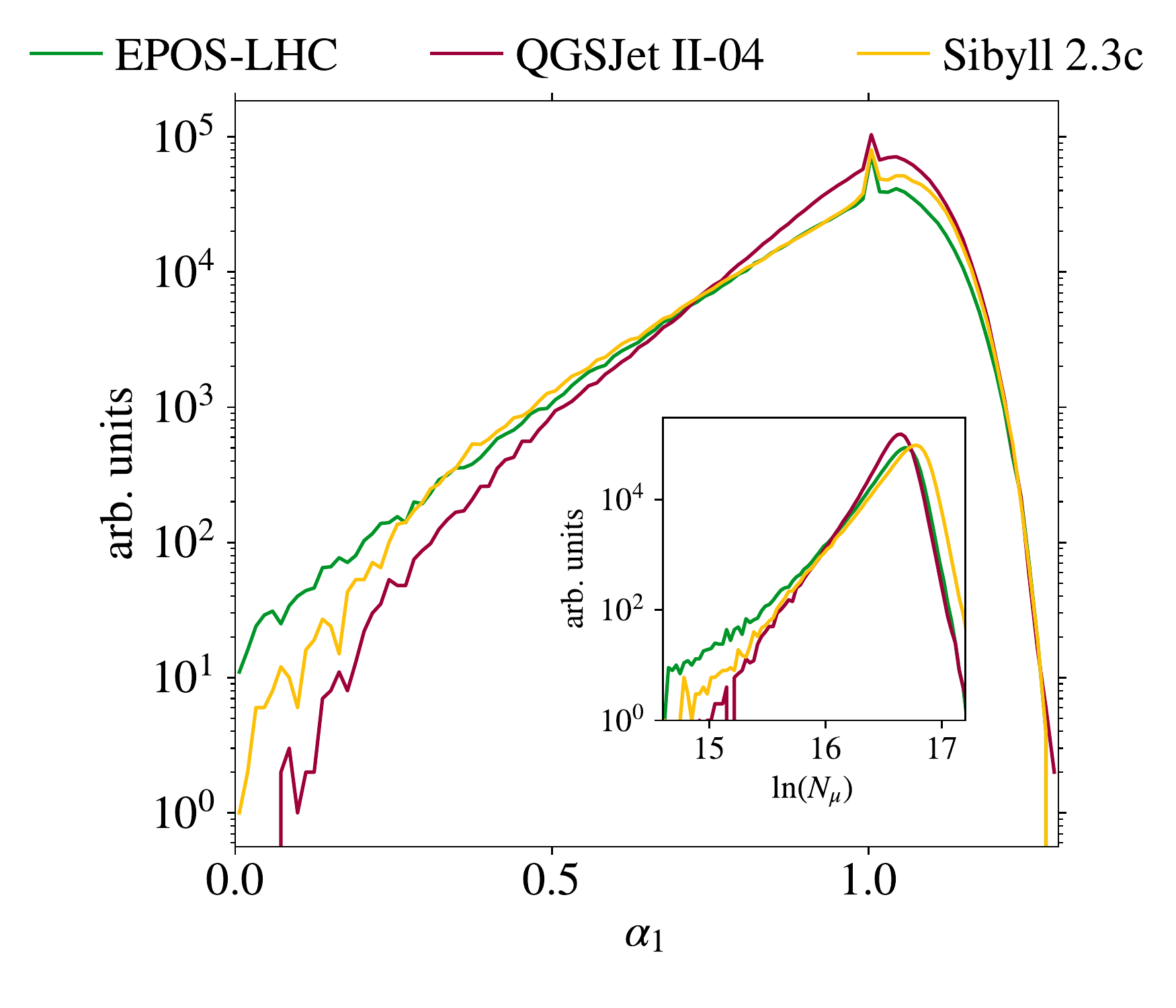}
  \caption{\label{fig:lalpha_lmu} Shower-to-shower distribution of $\alpha_1$ (large figure) and \LogNmu (inset figure) for proton-induced showers of $E_0=10^{19}\,$eV and $\theta=67^\circ$. The different lines are the predictions for distinct hadronic interaction models: \epos (\eposColor); \qgs (\qgsColor), and \sib (\sibColor). }
\end{figure}

Around one million showers were simulated with the primary energy $E_0 = 10^{19}\,$eV, at a zenith angle of $\theta = 67^\circ$ and the ground level at $1400\,$m above sea level, corresponding to the configuration under which the muon content in showers was measured at the Pierre Auger Observatory~\cite{AugerNIMA,AugerInclined}. High energy interactions were simulated with the post-LHC hadronic interaction models: \epos~\cite{Pierog:2013ria}, \qgs~\cite{Ostapchenko:2010vb}, and \sib~\cite{PhysRevD.102.063002}.

\begin{figure}[t!]
  \centering
  \includegraphics[width=\figsize\columnwidth]{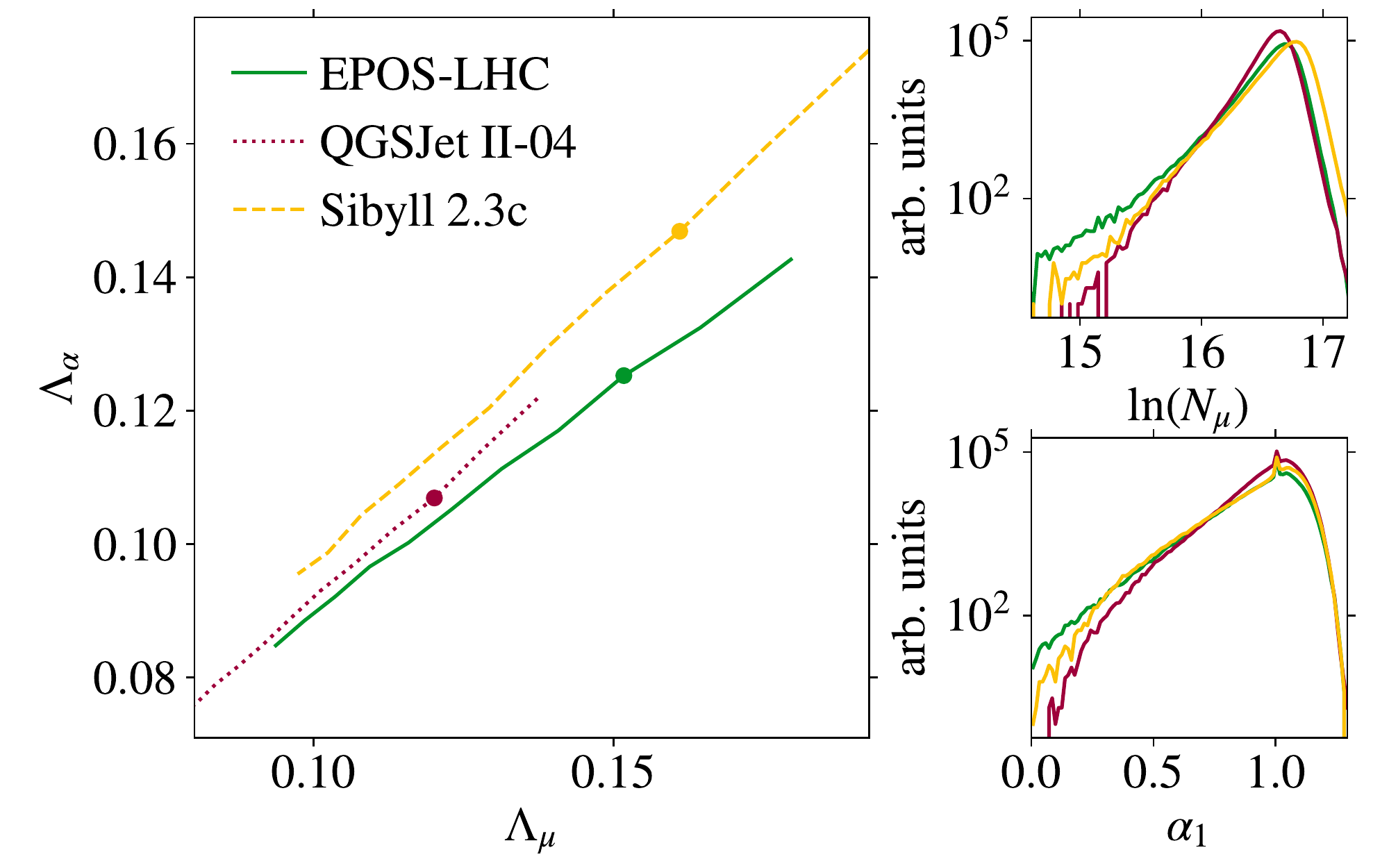}
  \caption{\label{fig:calib_lalpha_lmu} Conversion between \Lmu and \Lalpha. The filled circles indicate the predictions by the different interaction models. The lines show how \Lmu changes if \Lalpha is changed (see text for details).}

\end{figure}

In Figure~\ref{fig:lalpha_lmu}, the distribution of the natural logarithm of the muon content of showers, \LogNmu, and the distribution of $\alpha_1$ from the first interaction are shown for an ensemble of proton-induced showers. The peak at $\alpha_1=1$ corresponds to quasi-elastic scattering in the first interaction, which corresponds to a diffractive interaction between the cosmic ray and the atmospheric nuclei. The region of low values of both distributions was fitted to the functions $A \exp( \alpha_1/\Lambda_{\alpha})$ and $B \exp( \ln N_\mu/\Lambda_\mu)$, with $A$ and $\Lambda_{\alpha}$, and $B$ and $\Lambda_\mu$ as free parameters. The fitting ranges were chosen so that the deviation from a pure exponential function would not exceed 5\%~\footnote{This technique has the advantage of being more independent of the model and closer to what could be done in an experimental analysis. It was also checked that the qualitative behavior was insensitive to small variations of the fit range.}.
The values of \Lalpha and \Lmu for the three available models are represented in Figure~\ref{fig:calib_lalpha_lmu} by the solid dots. The relation between \Lalpha (related to the physics of the first interaction) and \Lmu (related to the muon content of the showers) was further studied. Small perturbations introduced in the $\alpha_1$ distribution of each model would result in a new value for $\ln N_\mu$. These perturbations were emulated by re-sampling the original data set and keeping pairs $(\Lambda_\mu,\Lambda_\alpha)$ with a probability $\propto \exp( \alpha_1/\delta \Lambda_{\alpha})$, where $\delta \Lambda_{\alpha}$ is the size of the perturbation. Note that in this way, the slope of the tail of the $\alpha_1$ was changed while preserving any non-exponential features. The new $\alpha_1$ and $\ln N_\mu$ distributions were fitted again to exponential functions, obtaining new values of \Lalpha and \Lmu.
The method used to perturb the $\alpha_1$ distributions works by directly altering the frequencies of the events and does not attempt to explain the fundamental physics causing these changes, which would also impact other basic parameters of the multiparticle production. 
A study of this kind is out of the scope of this paper, as this work concentrates on the experimental accessibility of the $\pi^0$ spectrum and $\alpha_1$ distribution, regardless of their deeper meaning, which must be investigated elsewhere. Note, though, that the three hadronic interaction models are instantiations of different physical laws, and one can observe that they scatter along the same general directions as the “frequency” perturbations. The same strategy has been used to measure the p-Air cross section in~\cite{Auger:2012wt}, where only a single parameter was modified while keeping the rest of multiparticle characteristics constant.

The result of this procedure is presented in Fig.~\ref{fig:calib_lalpha_lmu}. The points are the nominal values (\Lmu,\Lalpha) for each hadronic interaction model, and the lines are the results when the $\alpha_1$-distribution is varied with the resampling method. Notice that the resampling method assumes that the properties of the hadronic interaction remain unchanged. Hence, the conversion curves for each model cover only a limited region around the nominal model value.
Nevertheless, there is a monotonic relation between \Lmu and \Lalpha, where the maximum deviation between models in $\alpha_1$ is $\sim 19 \%$.
Through this conversion curve, one can transform a measurement of \Lmu into a quantity characteristic of the production cross section of the first interaction of the UHECR, \Lalpha.
It should be noted that the exponential tail in the distribution of \LogNmu is most prominent for proton-induced showers (largest \Lmu). For heavier primaries, the tail becomes steeper, (smaller \Lmu), such that their \LogNmu-distribution is hidden by the proton \LogNmu-distribution. While this means that the calibration between \Lmu and \Lalpha is only valid for proton primaries, it also means that the tail in the distribution of \LogNmu for protons can be measured even in the case of the scenario of a mixed mass composition. For that, it is only required that an enough number of proton showers is present. The effects of different mass composition scenarios will be discussed in Sec.~\ref{sec:exp}.

Each $\alpha_1$ value contains information about the energy distribution of the products arising from the first interaction, while \Lalpha gives information about the shower-to-shower distribution of $\alpha_1$ itself. In this sense, the latter is a more fundamental variable being more directly relatable to the hadronic interaction and the measurements performed in accelerators.

To see this, let us, for simplicity, consider pions as the only products of the hadronic interactions~\footnote{Pions constitute about 70\% of the particles arising from a high-energy interaction.}.
Due to the isospin symmetry, the number and energy carried by the bulk of neutral and charged pions emerging from the interaction should be equal. As such, the energy imbalance toward $\pi^0$, characteristic of showers with a reduced muon number, cannot be justified by low $\pi^{\pm}$ multiplicity, but instead, by fast $\pi^0$. These particles are, in essence, connected with valence quarks and forward production. Provided that the largest share of energy is taken by a single leading\footnote{the most energetic particle arising from an hadronic interaction is usually called the \emph{leading-particle}}  $\pi^0$, the remaining energy budget (shared among mesons and baryons) feeds the hadronic cascade, and it is partially used to produce muons.

\begin{figure}
  \centering
  \includegraphics[width=\figsize\columnwidth]{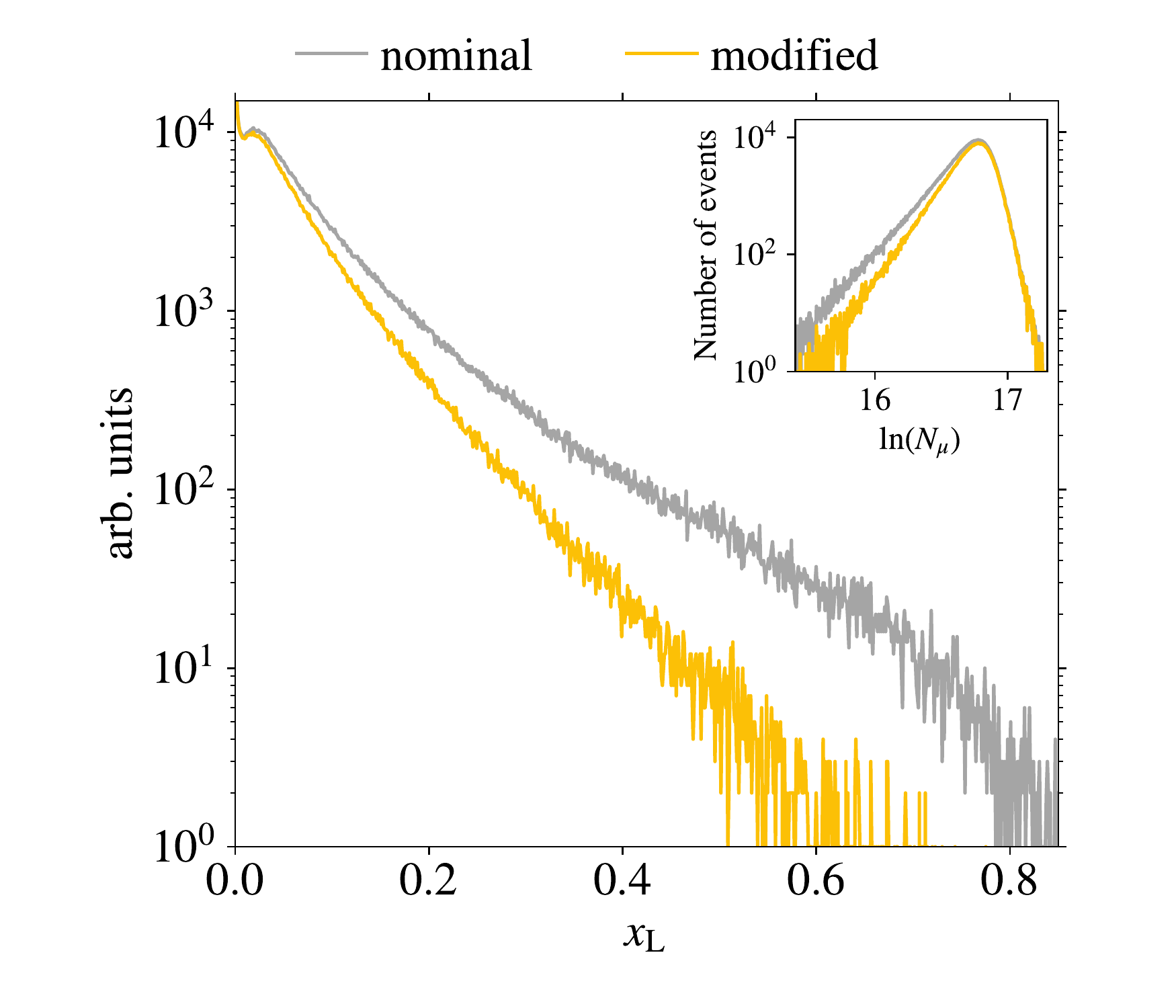}
  \caption{\label{fig:lmu_spectrum} Distribution of the fraction of energy in the laboratory frame carried by the highest energy $\pi^0$ in the first interaction of a proton with $E_0~=~10^{19}\,$eV. The gray distribution is the standard distribution in \sib while the yellow is for a modified neutral pion energy spectrum. The corresponding muon-number distributions for the air showers are shown in the inset plot.}
\end{figure}

Hence, it is only natural to investigate the connection between \Lmu and the high end of the $\pi^0$ energy spectrum. To do so, returning to the same shower simulations, the fraction of energy carried by the leading pion of the first interaction, $x_\textrm{L}$, was calculated for each event, giving the distribution shown in Fig.~\ref{fig:lmu_spectrum}. The tail of this distribution was also fitted to an exponential function of the form $C \exp(x_{\textrm{L}}/\Lambda_\pi)$, leaving $C$ and $\Lambda_\pi$ as free parameters. Once again, these simulations were resampled so that the nonexponential features of the distribution in the tail at large $x_\textrm{L}$ are preserved. As seen in the inset of Fig.~\ref{fig:lmu_spectrum}, this leads to a change in the tail of the muon number distribution with little effect to the rest of the distribution. Moreover, it can be seen that the harder the pion energy spectrum is, the smaller is the slope of the \LogNmu distribution, as there is less energy available to create muons.

\begin{figure}
  \centering
  \includegraphics[width=\figsize\columnwidth]{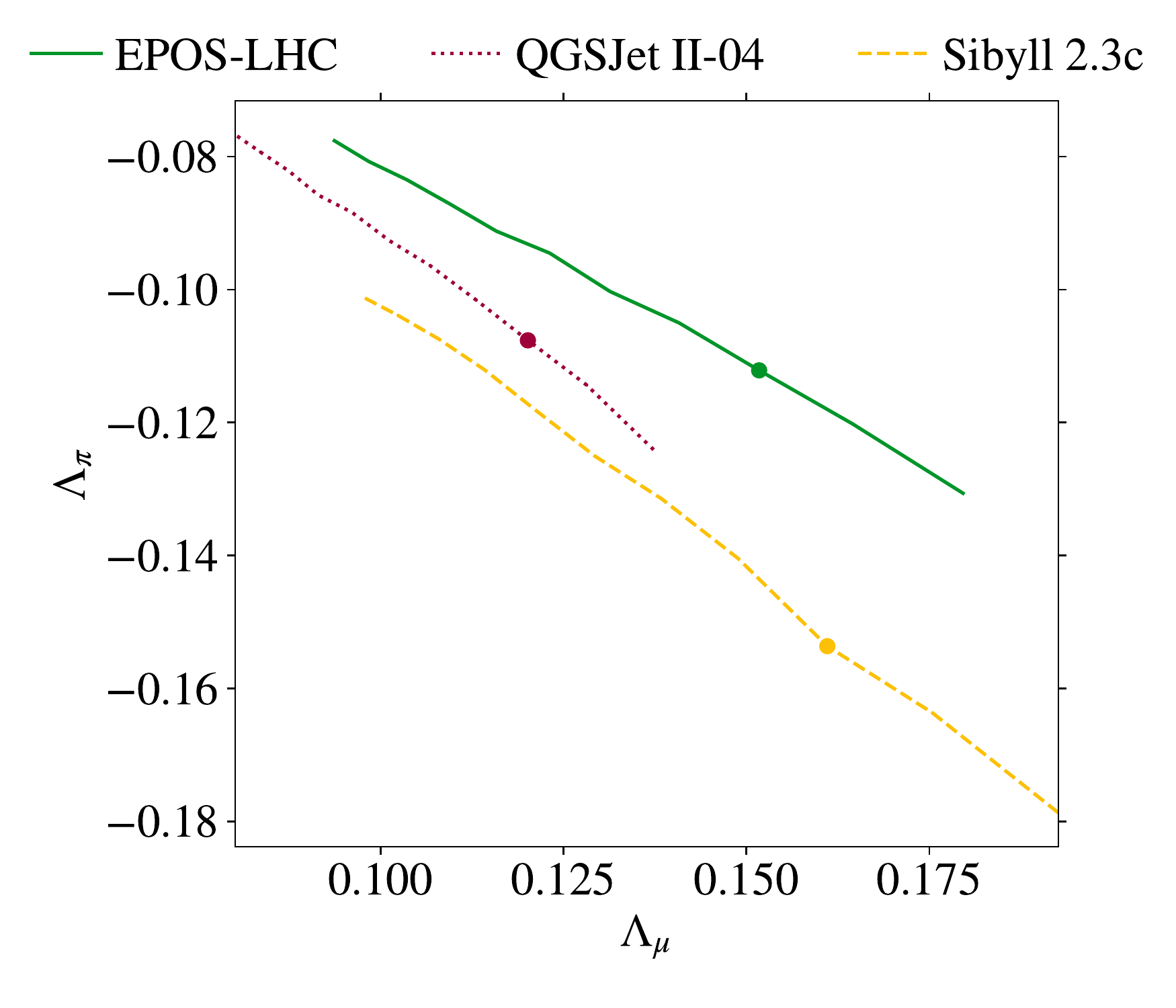}
  \caption{\label{fig:calib_lmu_lpi} Conversion between \Lmu and \Lpi. The filled circles indicate the predictions by the different interaction models. The lines show how \Lmu changes if \Lpi is changed (see text for details).}
\end{figure}

Following a similar strategy to the one performed in Sec.~\ref{sec:alpha} the relation between \Lmu and the slope of the pion energy spectrum high-energy tail, \Lpi, can be checked.
The conversion between \Lmu and the high energy tail of the $\pi^0$, \Lpi, is shown in Fig.~\ref{fig:calib_lmu_lpi} for each model. From this figure, it can be seen that this conversion is well defined under the interpretation of a given model. There is a remnant model dependence reaching up to  $\sim 35 \%$ difference. Despite this, provided that \Lmu can be experimentally accessed, this conversion could already be used to investigate possible exotic scenarios in the hadronic interactions at energy scales that surpass $100\,$TeV center of mass, which corresponds to $10^{18.7}\,$eV in the laboratory frame.

\section{\label{sec:exp}Measurement of \Lmu}

The previous results show the quantitative relationship between observables at the air shower level (\Lmu) and the production cross section of the highest energy $\pi^0$ in proton-air collisions (\Lpi).
In this section, we will discuss under what conditions it is feasible to measure \Lmu in proton-air interactions.

A realistic distribution of the muon content in showers with energy of $E_0 = 10^{18.7}\,$eV and zenith angle of $\theta = 67^\circ$ was built. Inclined showers are one of the few nearly-direct measurements of muons for showers initiated by UHECRs~\cite{AugerInclined}. Different primaries were used to emulate scenarios of a possible mixed primary composition~\cite{AugerMixedMass}, namely: proton, helium, nitrogen, iron, and photons. The experimental uncertainties were conservatively considered by smearing each entry in the histogram of muon numbers by $20\%$~\footnote{In case of the Pierre Auger Observatory, $\sigma^2~=~\sigma^2_{\rm res}(N_{\mu})+\sigma^2_{\rm res}(E)~=~0.15^2+0.08^2~=~0.17^2$~\cite{RMS:ICRC2019}}.

\begin{figure}
  \centering
  \includegraphics[width=\figsize\columnwidth]{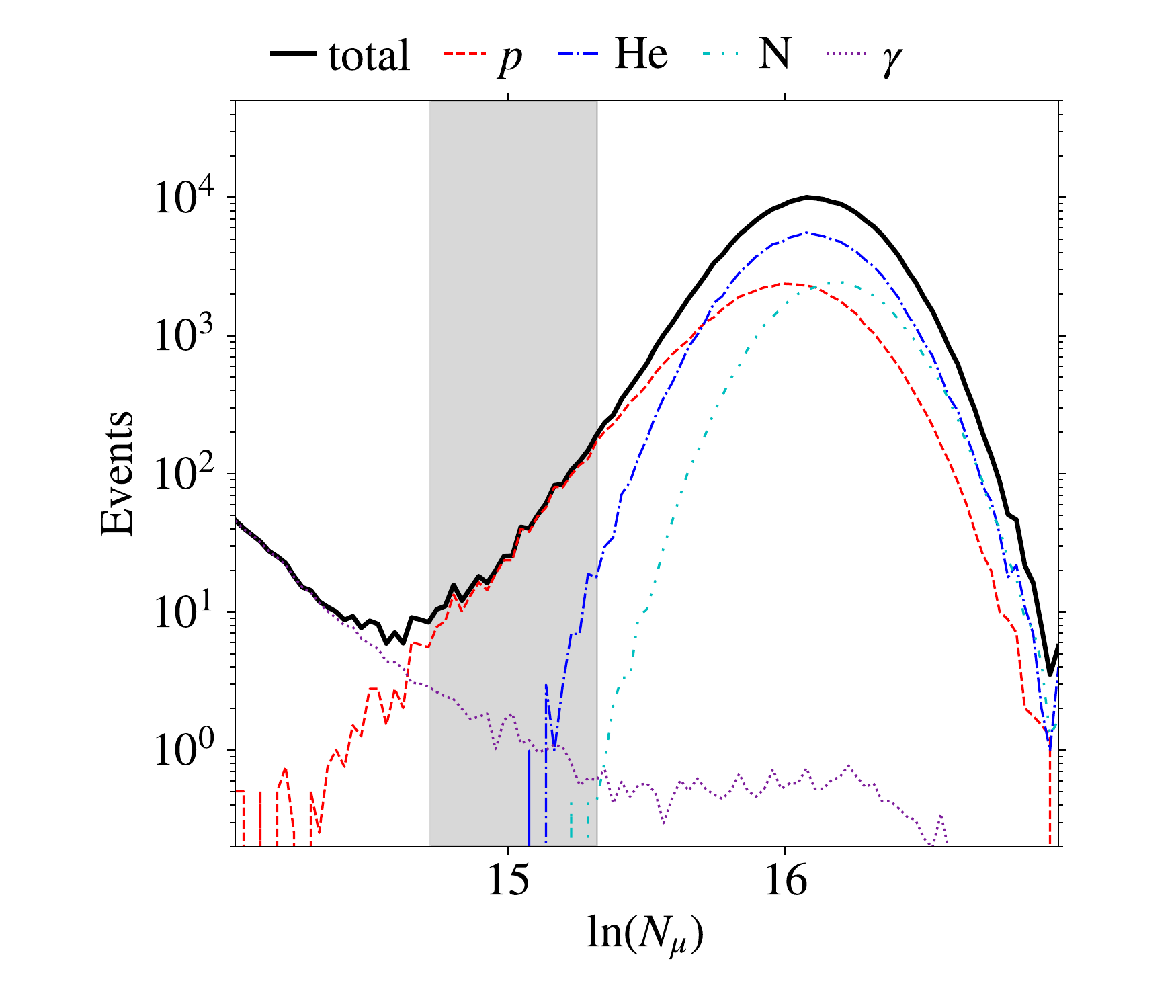}
  \caption{\label{fig:lognmu} Muon number distribution (see text for details on primary composition). The showers have been generated using \sib, the primary energy and zenith angle are $10^{18.7}\,$eV and $67^\circ$, respectively. A resolution of $20\%$ has been applied to the simulated muon number. The fit region chosen to extract \Lmu appears as a shaded band (see text for details).}
\end{figure}

The result of the above procedure for one of the considered mass composition scenarios can be seen in Fig.~\ref{fig:lognmu}. In this scenario, which is inspired by the mass composition from measurements of $X_{\textrm{max}}$ at these energies~\cite{AugerXmaxMass}, the ratio between p:He:N:Fe is taken to be 2:1:1:0. There has been no observation of ultra-high energy photons; so far, only upper limits have been determined by experiments~\cite{AugerPhotons}. However, pure electromagnetic showers produce a small number of muons at the ground, which can fall in the region where the low-\LogNmu tail is expected to be measured. As such, it was decided to include also the possible impact of photon contamination in the measurement of \Lmu. At the studied energies, the fraction of photons could be as high as $0.5\%$~\cite{AugerPhotons}.
A close inspection of Fig.~\ref{fig:lognmu} allows us to immediately see that the main sources of background in the measurement of the proton-induced showers \Lmu are photons and helium nuclei. Nevertheless, as shown in the same figure, these contributions can be minimized by selecting an appropriate region, as marked by the shaded band. The lower limit of the fit region removes bias on the \Lmu caused by photon induced showers while the upper limit cancels possible contributions from higher masses.

The overlap in Figure~\ref{fig:lognmu} between the total distribution and the distribution for protons in the shaded area is a good indication of the feasibility of the measurement within the current experimental conditions. Nevertheless, to make a more quantitative statement, the parameter \Lmu was extracted by fitting an exponential function to the tail of the total distribution. The obtained value was compared to the \Lmu value accessed from the proton distribution in the same fit region. In this way, the bias and the relative accuracy of the measurement could be determined. The later quantity is defined as $\delta_{\Lambda_\mu} \equiv 1 - (\Lambda_{N_\mu}^{\rm fit} / \Lambda_{N_\mu}^{\rm p})$, where $\Lambda^{\rm fit}$ and $\Lambda^{\rm p}$ are the exponential slopes of the tails of the total distribution and the distribution for protons, respectively. The bias and $\delta_{\Lambda_\mu}$ were analyzed for different hadronic interaction models, different primary mass composition scenarios, and the number of shower events necessary to accurately measure \Lmu.




\begin{table}[!h]
\caption{Table with the number of shower events in the fit region (tail) and in all the distribution (total) to perform the measurement of \Lmu with a precision of $\delta_{\Lambda_\mu} = 0.2$. The results are shown for the different hadronic interaction models and distinct mass composition scenarios (p:He:N:Fe).}
\label{tab:results}
\begin{tabular}{cccc}
\hline
Scenario                 & Model                & $N_{\rm evt}$ tail & $N_{\rm evt}$ total \\ \hline
\multirow{3}{*}{1:1:1:1} & \qgs  & 21      & 1564     \\
                         & \epos & 29      & 1926     \\
                         & \sib  & 30      & 1667     \\ \hline
\multirow{3}{*}{1:2:1:0} & \qgs  & 32      & 7086     \\
                         & \epos & 36      & 5505     \\
                         & \sib  & 33      & 4411     \\ \hline
\multirow{3}{*}{1:6:2:0} & \qgs  & 205     & 385776   \\
                         & \epos & 132     & 136212   \\
                         & \sib  & 123     & 78482    \\ \hline
\end{tabular}

\end{table}

Table~\ref{tab:results} summarizes the results obtained in this study. It was found that if the fit converges, then the bias on the measurement becomes negligible even in the most extreme scenarios. In this table,  the number of events necessary to reach an accuracy of 20\% on the measurement of \Lmu is also shown. The 20\% accuracy was chosen since it is the requirement to start distinguishing between hadronic interaction models. From these results, it becomes clear that, even in scenarios with an extreme mass composition, like the 1:6:2:0, it is still possible to measure the slope of the tail, provided that the number of events is large enough. This is because the exponential tail of the \LogNmu-distribution for helium falls off more quickly~\footnote{The \Nmu distribution becomes more Gaussian with the increase of the primary mass composition.}, which facilitates the measurement of \Lmu.
Given that the mean number of muons in a shower scales almost linearly with the shower energy, the energy spread of the data set translates directly into a spread in $N_\mu$. This effect can be mitigated by rescaling the muon number as $N_\mu / E_0^{\rm rec}$, where $E_0^{\rm rec}$ is the reconstructed primary energy. Provided that the experimental resolution on $E_0^{\rm rec}$ has a Gaussian form, $N_\mu / E_0^{\rm rec}$ does not hide the exponential tail.

The precise determination of \Lmu requires the measurement of \Nmu for around $3\, 000$ shower events, which is still a difficult number to reach~\cite{AugerInclined, TAmuons}. However, future experiments like GRANDProto300~\cite{GrandProto300:ICRC2019,Alvarez-Muniz:2018bhp} or upgrades, such as AugerPrime~\cite{AugerPrime}, where the muonic component is expected to be measured (directly or indirectly), may enable this measurement within a few years.
An important measurement of the number of muons is being conducted at the Pierre Auger Observatory using underground muon detectors~\cite{AMIGA} and, in the future, the number of muons can also be measured with resistive plate chambers~\cite{MARTA}. Both these extensions measure air showers in an energy range of about $E \simeq 10^{17}\,$eV, corresponding to the equivalent center-of-mass energy reached by the Large Hadron Collider. Therefore, the proposed measurement can be compared to measurements of accelerator experiments such as LHCf~\cite{PhysRevD.94.032007}, which are already measuring the forward energy flow carried by neutral pions in proton-proton and proton-lead collisions.

\section{\label{sec:conclusions}Conclusions}

The distribution of the muon number in air showers is sensitive to the properties of multiparticle production in the first interaction between the primary cosmic ray and the nuclei in the air. In typical scenarios with different UHECR masses, showers with the lowest muon number can be attributed to proton showers, which exhibit fluctuations to extremely low muon numbers. We find that the shape of the distribution at low muon numbers can be described by an exponential function with a characteristic slope \Lmu.
We show that \Lmu correlates with the multiparticle production property \Lalpha, which can therefore be determined at center-of-mass energies ranging from current accelerator possibilities to the $100\,$TeV scale through the measurement of \Lmu. 
We further show that there exists a more fundamental relation between the slope \Lmu and the slope of the production cross section of $\pi^0$ in proton-air interactions (\Lpi), 
 which can be constrained by the measurements of \Lmu as well.
Finally, it was shown that the measurement of \Lmu is already possible within the present experimental uncertainties, being only dependent on the event statistics, which shall be significantly improved by future upgrades.
\begin{acknowledgments}

The authors would like thank the colleagues from the Pierre~Auger~Collaboration for all the fruitful discussions. The authors would also like to give special thanks to S.~Andringa, L.~Apolin\'ario, G.~Parente, M.~Pimenta, D.~Schmidt, D.~Veberi\v{c} and J.~Vicha.

The authors thank also for the financial support by OE - Portugal, FCT, I. P., under project CERN/FIS-PAR/0023/2017.
R.~C.\ is grateful for the financial support by OE - Portugal, FCT, I. P., under DL57/2016/cP1330/cT0002. F.~R.\ acknowledges the financial support of Ministerio de Economia, Industria y Competitividad (FPA 2017-85114-P), Xunta de Galicia (ED431C 2017/07). This work is supported by the Mar\'\i a de Maeztu Units of Excellence program MDM-2016-0692 and the Spanish Research State Agency. This work is co-funded by the European Regional Development Fund (ERDF/FEDER program).
\end{acknowledgments}


%

\end{document}